# Role of Massive Radiation Pressure in the Early Universe
General Description
Andrzej Dworaczek avo@g.pl

**Abstract**

**The characteristic distribution of the matter in the universe and its observable expansion (most galaxies and their clusters are moving away from each other with apparent\* acceleration) is the result of historical event - massive stellar explosions' radiation pressure on drifting particles of hydrogen environment.**

*This process has began just after the first stars were formed from archaic giant dens Hydrogen nebulae devoid of any other matter (galaxies, stars, planets or even dust)*
*As a currently observed results of a radiation pressure is an apparently\* accelerating in its expansion universe and a characteristic, sponge-like distribution of matter in space.*
*The process described above is a consequence of the single event - cosmic inflation, or possibly - as a part of a repeating cycle.*

*Supporting thesis in short essays: "Playing with metal balls" ,"Broken Entropy" and "Single Particle Evaporation"*

**Introduction**

A few years ago, American scientists started a project to measure the distances to thousands of galaxies and in the process, developed a sliced flat map of our nearby universe showing a matter distribution in space - the 2dF Galaxy Redshift Survey project. Their map shows that the cosmos is arranged in a most intriguing way. Galactic clusters are separated by great distances mostly devoid of matter. The intriguing part is that the shape of this empty space appears to be spherical. Until this time scientists have expected to see a much different picture of the universe.

Why did our universe form this way? One of the most popular explanations of this distribution is the Cosmic String Theory, which tries to reconcile knowledge found in quantum mechanics and astrophysics. I am certain that the distribution of matter in our universe might have followed a different path and invite you to follow my train of thought, which lead me to different conclusions, and finally ends conceiving our universe in a unique way.

The cosmos is fascinating. Many questions, to which modern science is unable to provide satisfactory answers, unleash an avalanche of speculations. Astrophysicists take a different stance when it comes to the creation of our universe. During a time of examination, an array of theories arises, changes with time, and then disappears. The scientific evidences provided by ever evolving instruments surprise and, at times, contradicts common sense, further disproving those initial theories. Astrophysicists often speak of creating "Cosmological Models". This is most commonly understood as development of a simplified mathematical model that defines the history and structure of the universe; thus capturing its general properties. This type of a model can not explain all the characteristics of our universe.

"Cosmological Models" define the universe as though it was a flat sea of matter from which cosmos was built. In those theories a gathering of matter into stars and galaxies is completely ignored. The separation from homogeneous distribution is examined only when more detailed questions arise, (i.e. the origin of stars and galaxies). According to astrophysicists, this approach still gives surprisingly accurate results. During observations it is possible to find certain parameters of the universe, such as a density of matter in the universe and a temperature. If a model allows for such parameters it can be considered as a good description of the real universe.

**Elusively-hydrogen state of universe**

Let us imagine a following case:
In accordance with assumptions, a universe at its birth is expanding (or, on the contrary gaining density) without interruptions. From this follows, a uniform distribution of matter (mostly, or rather only a hydrogen), the temperature of which rich ~ 3K - not due to decompression, but rather rise high ~ 3K due to very slow gravitational force (compression). ( ~ 3K may represent a certain density of hydrogen as well, and may provide some information about the size of such dens hydrogen-universe). There were no groups of matter such as stars or galaxies. There was no expansion process yet. Let us assume that this is the case, then the average distance between particles statistically is about equal. The Universe has become just a huge dens hydrogen cloud, much smaller in size than the current Universe. Assuming that, at this point, Cosmic Inflation may be correct as well (as the cause of such state).

Thanks to a balance between gravity and expansion it has achieved some kind of stable state for a while. In other words, we are left with an unaltered universe; one that does not resemble the picture seen today. This state would be welcomed by cosmologists, because it is very predictable.
Regardless of uniform distribution of hydrogen particles in space, however small their movements ,will cause the formation of local concentrations due to loss of fragile gravitational balance. 3K~ of cosmic background radiation tells us that probably, as described above, the (huge hydrogen cloud)-universe was very dense, so Local concentrations will lead, with no doubt, to formation of first star(s).***

---

*** *Similar formations are now observable in stellar nurseries where emitted radiation-energy sweeps out its surroundings from particles, however in the currently observed universe the local quantities, dynamics and density of stellar gas cannot produce a lot of heavy stars causing stellar explosions - (supernova type).*

## Massive stellar explosions in the hydrogen-state universe

In the past, the results of such slow concentrations would lead to heavy star formations, heavy enough to cause massive stellar explosions and then enormous light (radiation) pressure. This is sweeping and consequently pressing particles process is accompanied by velocity (and momentum) of further spherical (expanding) concentrations of environmental hydrogen, that give way to easy formation of similar heavy stars consequently moving away together with their environment-(mostly spherical in shape distribution of surrounding matter). See illustration #2

Such occurrences accompanied the formation of the first stars in our universe. As I mentioned - the physical conditions of the primordial clouds were different from stellar clouds observed today.

Therefore, prior to formation of concentrations of (unlike currently observable in stellar clouds) particles that were under the influence of strong and massive exploding star radiation pressure, had the ability to travel much greater cosmic distances in a long period of time, producing locally almost flat (surface of huge spheres), dens formations and adding constant momentum to the next and consequently newly formed spheres - shaping slowly our expanding and apparently* accelerating universe, sweeping huge voids between thin clusters of new galaxies) This process had a strong influence to present galaxies shapes, their distribution, size and motion as well.

**Summary:**

**The characteristic distribution of observable matter in the universe and apparent* acceleration in expansion of the universe is the result of light pressure****

 ***This process has begun just after first stars were formed from archaic giant dens hydrogen nebulae devoid of any other matter (galaxies, stars, planets or even dust)*

In this context the **light** pressure is understood to be representative of all electromagnetic radiation emitting by any source of electromagnetic energy, i.e. star upon its environment throughout all the stages of a its evolution.
I used the term "for the most part", as it is important to keep in mind that the minor role played by gravity and eruption of matter during explosion of a certain type of heavy stars, however it was not significant in process of shaping characteristic distribution of matter (galaxies and their clusters locations) in the universe.

**Notes:**

I believe that the main role in creating the "Cosmic Deserts" was played by an another phenomenon. In the early stages of universe evolution, (due to very slowly thickening hydrogen) there most likely existed an environment more akin to formation of extraordinary massive stars – never observed by astronomers today (and commonly, massive spherical in shape galaxies/or other matter dens formations) in the early universe. For this type of starts (hundreds times heavier than Sun), beside of producing (in their final point of evolution) a huge luminosity and a correspondingly a large radiation pressure force on the surrounding hydrogen - from stellar explosion some heavy dust particles thrown at great speeds also act as sort of a gravitational glue to their surroundings, increasing the intensity of this phenomenon. Studying such cases we should take into consideration, that today a gravitational forces are different in value in dens formation i.e. galaxies.

The role of light pressure (radiation pressure) has been recognized some time ago. It is a proven tool in explaining many observed phenomena e.g. directions of comet tails. The creation of first stars must have occurred identically to the way stars are currently born in observed stellar nurseries, however, the particles' densities in space, their spin in space and gravitational environment of surrounding were different. The definite properties of condensing and shaping hydrogen clouds is the reason for smaller currently observed vacancies surrounding young stars in stellar clouds. The light's pressure works in stellar nursery is some way related to the early stage of the universe, however on a much smaller scale.

It is difficult for me to reconcile the theoretical existence of stars and galaxies in a much earlier stage of universe development (cosmic inflation models showing young galaxies). However Ideal Inflation models of universe expansion happily predict even distribution of Only hydrogen particles in the universe. Thus, lets assume that at the beginning there were no stars and galaxies.

The distribution was even, as it is assumed in the simplest models. Such a distribution must sooner or later would lead to gravitational aberrations; meaning that drifting particles were unable to maintain equal distances between each other. In other words: maintaining a balance between particles for a prolonged period of time is very unlikely as even the slightest change will disrupt this balance. Created in this way, particles' groups gave rise to first and dens stellar clouds and thus creation of first (because of characteristic particles' slow spherical concentrations with very low spin - rather very heavy) starts.

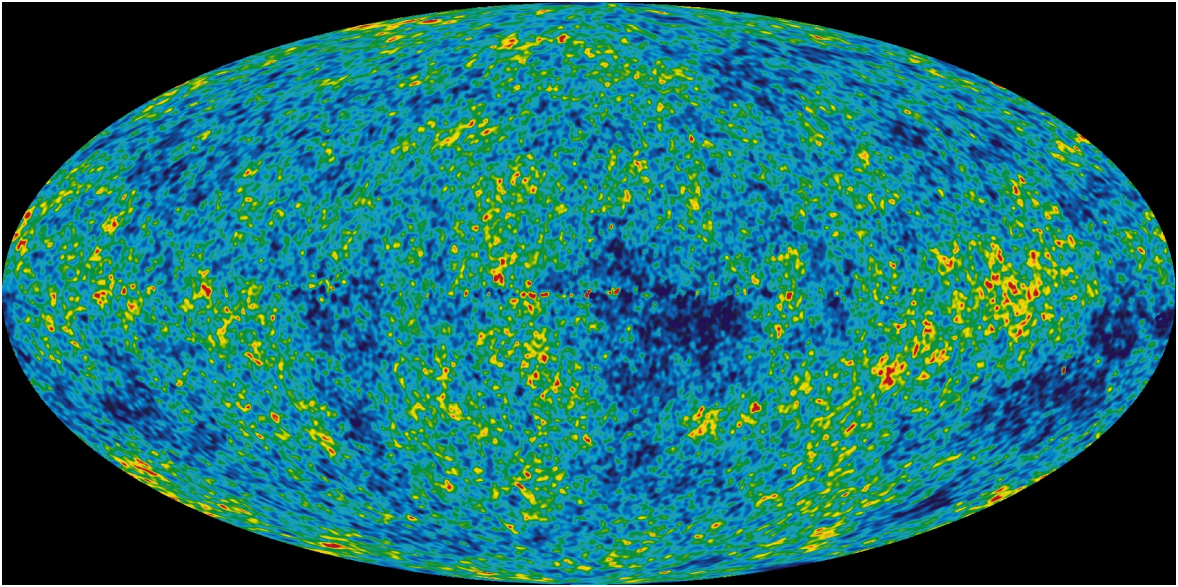

*Credit: NASA/WMAP Science Team*
*The illustration #1 shows the microwave background radiation. The minimal deviations of temperature (colors) are parallel to the expected hydrogen super-cloud density deviations – consequences are described above.*

    Please take notice of the emitted radiation distribution. In my opinion, it clearly shows the creation of the further particles concentrations and their vacancies.
    The effect of light emissions pressure, observable today in the stellar nurseries, surrounding young stars, is the empty volumes devoid of any particles. Sphere of gas (with some matter dust) constantly moving away from a star provides a supporting environment for birth of new stars (witch are moving away as well – as they belong to their environments) The event is repeating, (but not fractal). The process described above should create a universe with matter distributed in the manner in which it is currently observed.
    Additionally, a final point of this processes should concern the empty spherical areas of space (voids) named in my thesis - Cosmic Deserts.
    A different phenomenon accompanying previously described process is the "re-expanding" matter with acceleration. (evolution of velocities)

### Evolution of velocities

    Let us assume that before (described above) expansion of the universe, achieved a balanced state, or rather began to contract (gain density). So, the gravity equaled the velocity given to particles in the primary stage (assuming the theory of expansion "Big Bang" is compliant with currently available data - if such ever happened). As the process of emission-particle expansion (i.e. light venturing further into space) accompanying the birth of first stars must have begun locally, the examination of forces vectors for changes in time must show acceleration of expansion. There is only one conclusion: universe should expand at ever increasing speed until galaxies are shaped.

The expected evidence of my theory and courage to speak about it, was provided for me in 1999 through the results (published in 1998) of experiments conducted by Samuel Perlmutter group - "Supernova Cosmology Project" at Lawrence Berkeley National Laboratory.

Why is it easy to ignore the radiation pressure in our mind?
Often, when examining forces working in dense particle fields, such as planets or galaxies, radiation pressure is ignored.  This omission is understood due to following reason: stars and planes are subjects to immense gravitational fields.  The effect of light pressure is minuscule when compared with the condensed mass of those solar bodies (Radiation pressure is inconsequential for large bodies, but it has a significant effect on exposed small particles, especially on single drifting light hydrogen particles).  The conditions perceived at planetary scale can be thought of completely differently.

The effect of gravity in hydrogen-cloud state can be set aside as it acts almost equally, regardless of direction,  on every particle.  If such drifting particles affected by omnidirectional and balanced force of gravity, then any directional force (i.e. star light pressure) constantly applied to a particle will result in increased its velocity in time.  The amount of change is not relevant here; a child may move a train - if friction is removed.  If a child kept applying that small force for billions of years this train would achieve an unimaginable velocity.

An excellent example of concept is the Crab Nebula where the influence of radiation pressure, years after the explosion of a supernova, is clearly visible.  After merely a thousand years from it's explosion, the discarded particles span an area of six light years.  After an estimation of mass was completed, it was concluded that some particles were from the exploded star.  The remaining particles was gas surrounding that star before stellar explosion.

I assume that by observing of incoming light from far space it would be much easier way to find massive black holes left by first heavy stars and their archaic formations.

First and primitive "galaxies" (source of enormous radiation in past) must have had very low rotation (spin) against the rest of the "Super Nebulae-Universe" and were rather spherical due to slow, equal gravity force, so they finally must have rather easy collapsed into extremely heavy black holes, which location are expected to be near by the center of each observable empty spherical space separating groups of galaxies and galaxy clusters.

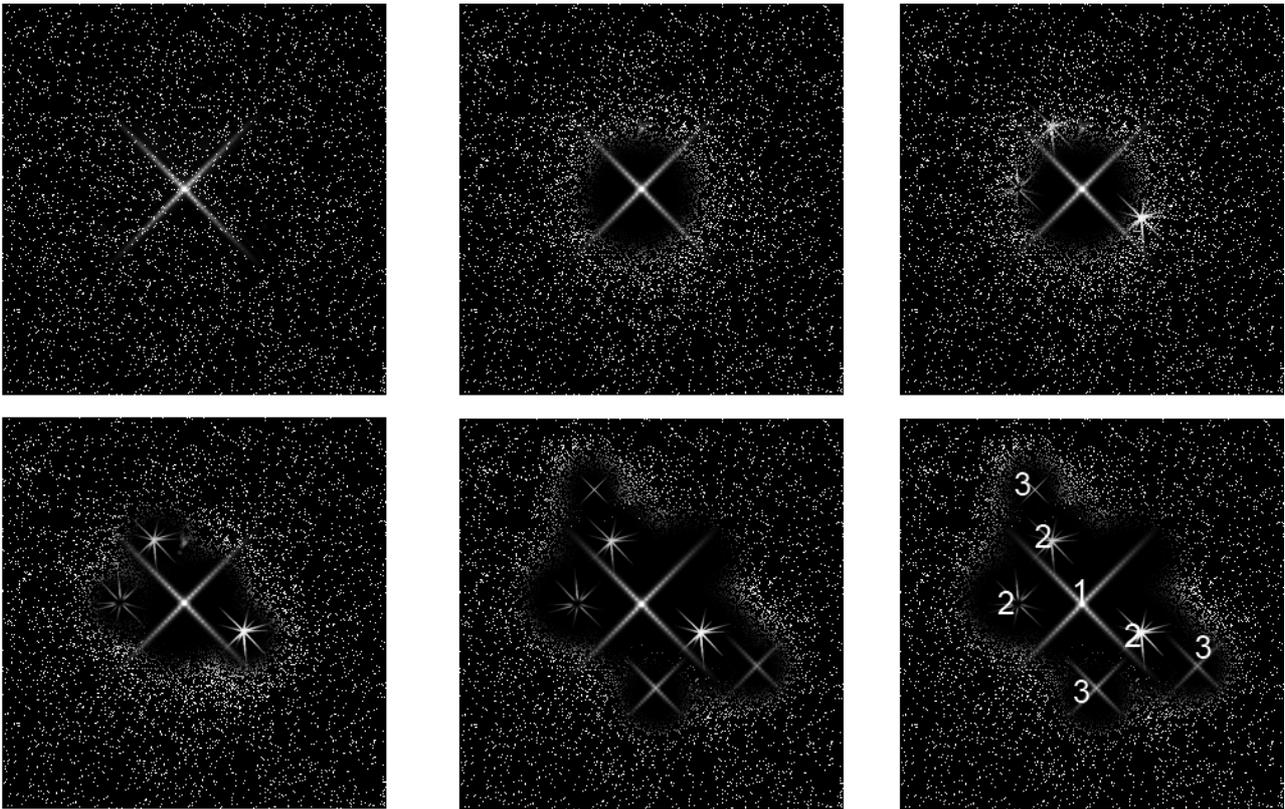

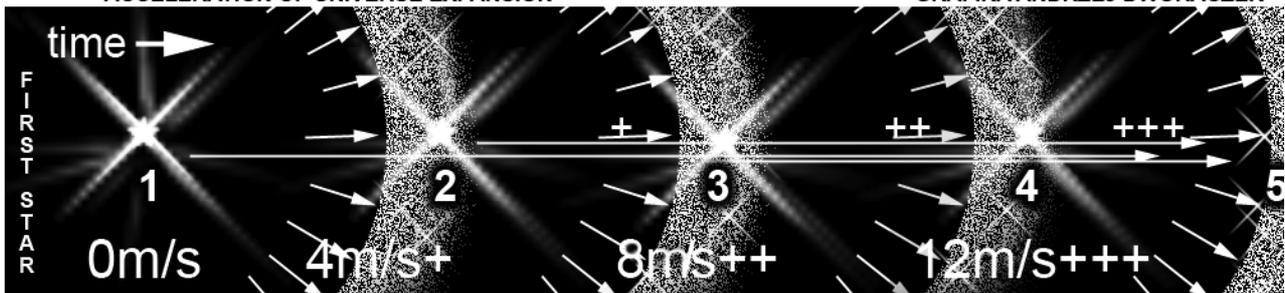

The illustration #2 shows the speed development in the early universe (speed numbers on the bottom of picture from 0m/s to 12m/s+++ are only examples (not actual speed) - I used them to make the acceleration steps of expansion process caused by light pressure easy to understand.

*I was looking for some expected behaviors and evidences, in the Universe and among others on some pictures from space telescopes.*
*As the effect of my effort - there are:*
- *thin strands of galaxy clusters around the huge bubbles of empty space*
- *expected Colliding Galaxies - to constitute sufficient evidence of behaviors of matter in space and time - in expectation of galaxy mergers, - I used a primitive simulating tool – I observed growing yeast between thin glass plates. Their gas bubbles did not grow equally in all directions (size, velocity). In Conclusion it is necessary for neighboring galaxies to have some different movements against each other, like galaxy mergers (some various velocity and aberration in directions)*
- *illusory\* acceleration of space expansion – described below in an essay \*"Playing with metal balls"*
- *present distribution of Spherical Galaxies in the universe*
- *shapes of Galaxies*

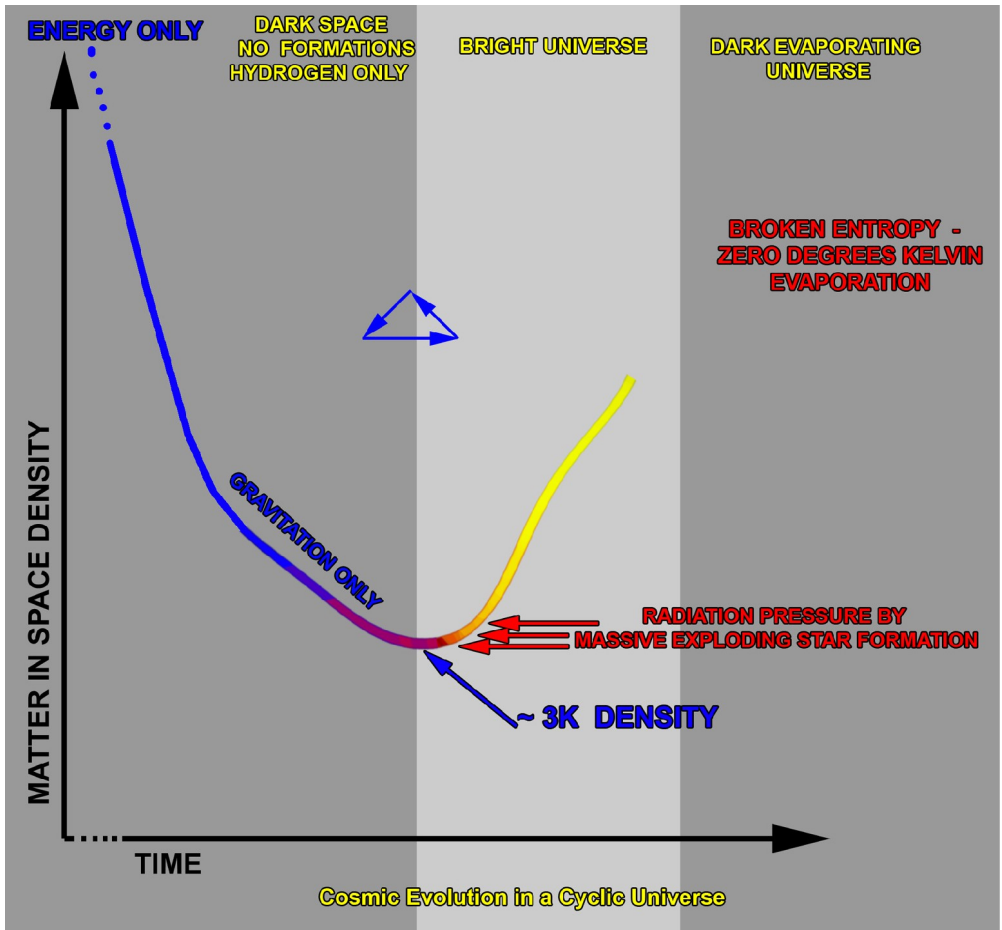
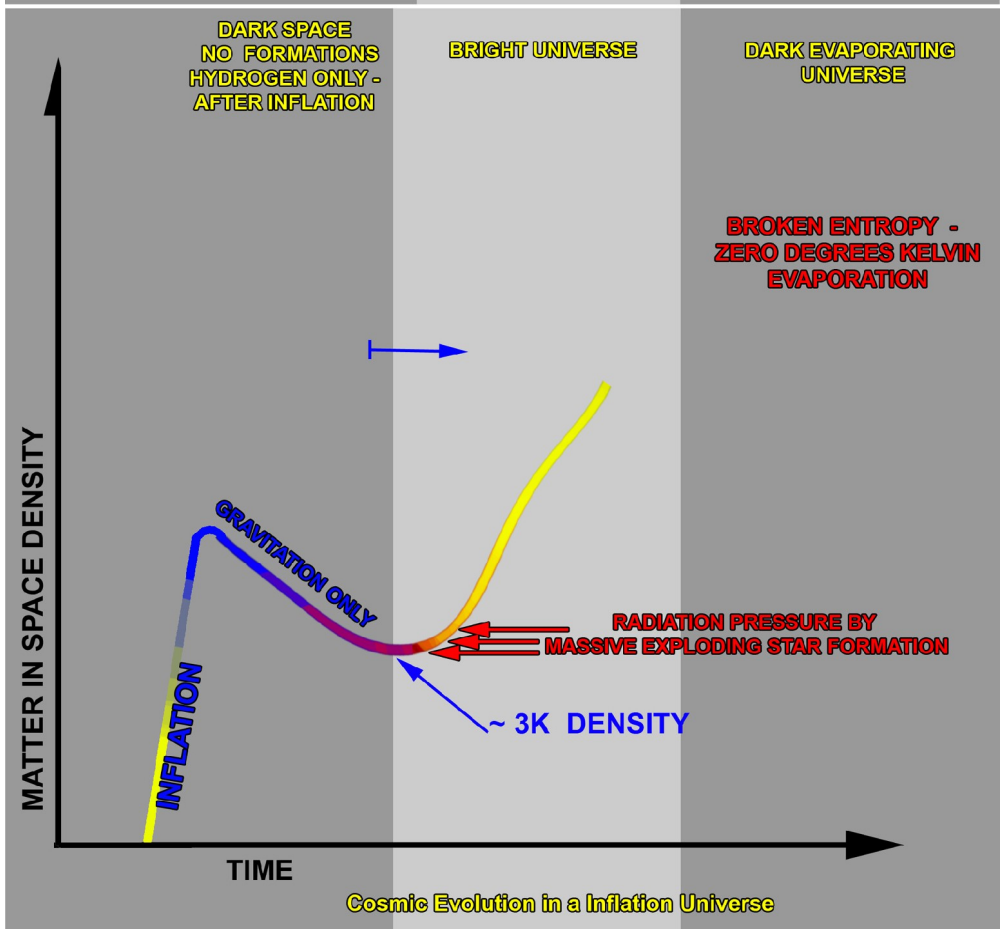

*Illustration #3 Matter in Space density (mass to size of the universe)*

# Supporting Thesis

## "Playing with metal balls"

Analyzing the spectrum of light (redshift) of stellar explosion - based on type Ia supernova - we might come to a conclusion, that the universe appears to be expanding at an increasing rate – is **accelerating -** "Supernova Cosmology Project"

However, this supposition might not be correct.

We just discover only, that the velocity of expanding universe is increasing proportionally to the distance from the observer. **(increasing velocity proportionally to the distance)**

Is it not the same? Answer is - No.

Explanation: The Source of Acceleration might not exist anymore (in other words: was expanding with a different rate vs - is expanding at an increasing rate).

Lets do an experiment:

Allow a few shiny metal balls roll down a decline, one after another in a constant period of time; for example one every two seconds.

When we measure the velocity in a very short period of time, we noticed that the velocity of ball number one is greater than that of number two and the second is rolling faster than the third etc.

After releasing the last ball we immediately turn off the light and **flatten** the slope. Nobody knows about it.

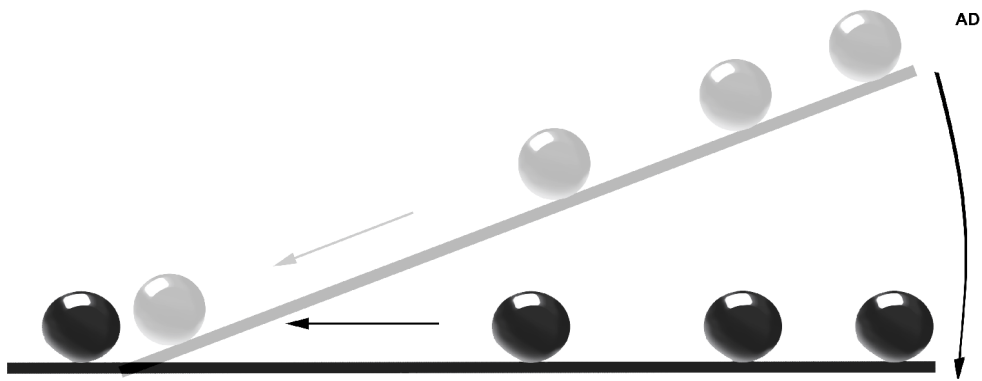

The balls are still rolling, but now on a flat, horizontal surface – so gravity force is not working any more. Each ball still has a different velocity, because gravitational force worked in a different period of time for each and each ball has different momentum.

All the balls are identical, and they eventually should reflect the same amount of light, from the white ceiling if the light is on*.

Someone, who doesn't know all the conditions of our experiment came to the totally dark room with super sensitive spectrograph, to check each ball velocity and position (by analyzing further pictures of light spectrum characteristics).

He or She has only one chance to measure reflected light coming from balls' shiny surface, when light was on – in other words: to measure speed of all balls together in one very short period of time. When the light was on He or She noticed that all balls are rolling on a flat horizontal surface.

He or She came left dark room to his/her own laboratory to analyze the results of incoming (from balls) light measurement.

The final analysis of spectrogram shows undoubtedly, that each ball has an unexpected different velocity - indicates acceleration.

The result is surprising, because he or she does not observe any reason for a such behavior.

Observed balls velocity is increasing proportionally to the distance from observer showing an acceleration of "balls' little universe".

Their conclusion is: "something is attracting balls", and next one is: "we do not observe anything, they come to the conclusion: "maybe is a magnet hidden in the opposite wall attracting balls?" maybe this is a reason of velocity differences?", but based on further mathematical estimations that conclusion seems to be not fully correct, so they add next one: "there must be some kind of invisible energy"

None of those conclusions are correct.

Similar experiment might be done in real world.

In place of shiny balls we can observe, and measure extremely luminous burst of light coming from distant supernovas' explosions – (we choose few type Ia for analyzing spectral records)

Base on the results we may estimate the distance to the exploding stars and their velocity. All basics of these experiments are the same (certain light energy, its' redshift and short time of measurement)****

---

****(*more accurate experiment basics related to „Supernova Cosmology Project" would be, if each ball is a source of one short flash of light, size of darkroom would be thousands of parsecs and balls have high speed, however foundations of experiment to describe ways of thinking are correct*)

# "Broken Entropy"

**The Entropy in a Real universe.**

The universe will never fate to a heat death.
Black holes are able to absorb energy - including thermal (or any other electromagnetic) radiation.
So finally the universe temperature is decreasing to zero degrees in Kelvin scale (absolute zero).
Below is an illustration of a perfect (lossless - in terms of kinetic energy ,friction etc.) billiard table . Total kinetic energy of all balls goes down in function of time, because sooner or later the last ball must fall into the hole.

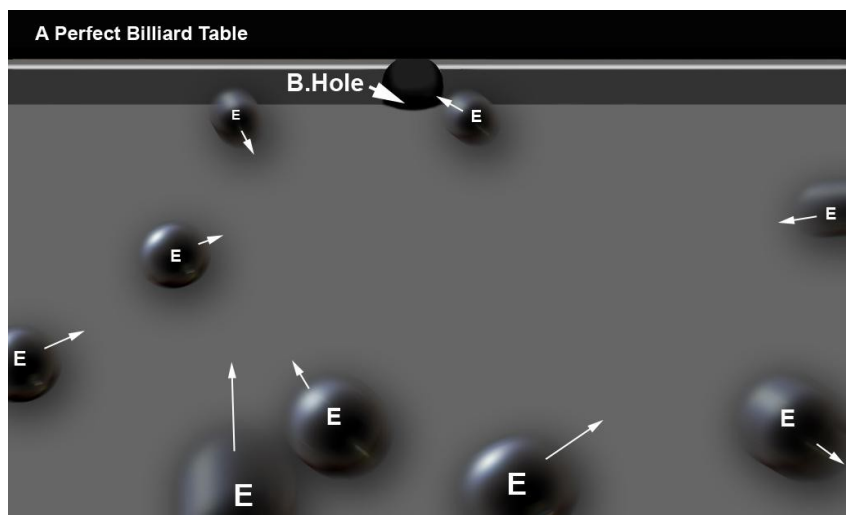

# "Single Particle Evaporation"

**The Energy State of the Universe**

Even if some particle's formations still will exist after the temperature of universe dropped to zero degrees Kelvin see: "Broken Entropy" above -  they must evaporate.
Explanation of evaporation process:
even if particles temperature dropped to 0 Kelvin, they are still emitting gravitational radiation, or any information (no information – no existence).
However there must be a source of energy for such emission inside such a particle.
    Emission costs a particle loss of its mass – (even if is a question about particles state of stability – quantum physics)
In a real universe particles mass is decreasing - until the last one fully evaporates.
This process includes black holes as well.
The universe will change to it's energy state (without a mass).